\documentclass[10pt,aps,prx,twocolumn,notitlepage,showpacs,superscriptaddress]{revtex4-1}
\usepackage{amsthm,amsmath,amsfonts,amssymb,verbatim, color}
\usepackage{graphicx}
\usepackage{bm,bbold}
\usepackage{epsfig,slashed}
\usepackage[colorlinks=true,citecolor=blue,
linkcolor=blue,urlcolor=blue]{hyperref}
\usepackage[caption=false]{subfig}
\usepackage[normalem]{ulem}
\begin{document}

\def\ie{{\it i.e.},\ }
\def\eg{{\it e.g.}\ }
\def\etal{{\it et al.}}
\def\etc{\emph{etc.}\ }
\def\ibid{\emph{ibid.} }
\def\cf{\emph{cf.}\ }
\newcommand{\tr}{\mathop{\mathrm{Tr}}}
\newcommand{\bsigma}{\boldsymbol{\sigma}}
\newcommand{\re}{\mathop{\mathrm{Re}}}
\newcommand{\im}{\mathop{\mathrm{Im}}}
\newcommand{\diag}{\mathrm{diag}}
\newcommand{\sign}{\mathrm{sign}}
\newcommand{\sgn}{\mathop{\mathrm{sgn}}}
\renewcommand{\c}[1]{\mathcal{#1}}
\newcommand{\bra}[1]{\left\langle#1\right|}
\newcommand{\ket}[1]{\left|#1\right\rangle}
\newcommand{\bigket}[1]{\bigl|#1\bigr\rangle}
\newcommand{\textket}[1]{|#1\rangle}
\newcommand{\vac}{\left|0\right\rangle}

\newcommand{\mb}{\bm}
\newcommand{\ua}{\uparrow}
\newcommand{\da}{\downarrow}
\newcommand{\ra}{\rightarrow}
\newcommand{\la}{\leftarrow}
\newcommand{\mc}{\mathcal}
\newcommand{\bs}{\boldsymbol}
\newcommand{\lra}{\leftrightarrow}
\newcommand{\nn}{\nonumber}
\newcommand{\half}{{\textstyle{\frac{1}{2}}}}
\newcommand{\mf}{\mathfrak}
\newcommand{\MF}{\text{MF}}
\newcommand{\IR}{\text{IR}}
\newcommand{\UV}{\text{UV}}
\newcommand{\bl}{\text{b}}
\newcommand{\g}{\text{g}}
\renewcommand{\r}{\text{r}}

\renewcommand{\a}{\alpha}
\renewcommand{\b}{\beta}
\newcommand {\ea}{\eta_{\alpha}}
\newcommand {\eb}{\eta_{\beta}}
\newcommand {\eab}{\bar\eta_{\alpha}}
\newcommand {\ebb}{\bar\eta_{\beta}}
\newcommand {\Sa}{S_{\alpha}}
\newcommand {\Sb}{S_{\beta}}
\newcommand {\Stot}{S_{\text{tot}}}
\newcommand {\bsS}{\bs{S}}
\newcommand {\bSa}{\bs{S}_{\alpha}}
\newcommand {\bSb}{\bs{S}_{\beta}}
\newcommand {\bSc}{\bs{S}_{\gamma}}
\newcommand {\zb}{{\bar z}}
\newcommand {\wb}{{\bar w}}
\newcommand{\A}{\text{A}}
\newcommand{\B}{\text{B}}

\DeclareGraphicsExtensions{.png}

\title{Universal entanglement spectra in critical spin chains}

\author{Rex Lundgren}
\author{Jonathan Blair}
\author{Pontus Laurell}
\affiliation{Department of Physics, The University of Texas at Austin, Austin, Texas 78712, USA}
\author{Nicolas Regnault}
\affiliation{Department of Physics, Princeton University, Princeton NJ 08544, USA}
\affiliation{Laboratoire Pierre Aigrain, Ecole Normale Sup\'erieure-PSL Research University, CNRS, Universit\'e Pierre et Marie Curie-Sorbonne Universit\'es, Universit\'e Paris Diderot-Sorbonne Paris Cit\'e, 24 rue Lhomond, 75231 Paris Cedex 05, France}
\author{Gregory A. Fiete}
\affiliation{Department of Physics, The University of Texas at Austin, Austin, Texas 78712, USA}
\author{Martin Greiter}
\author{Ronny Thomale}
\affiliation{Institute for Theoretical Physics, University of W\"{u}rzburg, D-97074 W\"urzburg, Germany}
\date\today

\begin{abstract} 
%We advocate that in critical spin chains {\color{blue}whose low energy
 % properties are described by an SU(N)$_k$ Wess-Zumino-Witten model}
%\sout{, and possibly in 1D critical models in general,} a gap in the
%\sout{momentum cut} {\color{blue}momentum-space} entanglement spectrum
%separates the universal part of the spectrum, which is determined by
%the associated conformal field theory, from the non-universal part,
%which is specific to the model. 
We advocate that in critical spin chains, and possibly in a larger
class of 1D critical models, a gap in the
momentum-space entanglement spectrum
separates the universal part of the spectrum, which is determined by
the associated conformal field theory, from the non-universal part,
which is specific to the model.  
To this end, we provide affirmative evidence from multicritical spin chains with low energy
sectors described by the SU(2)$_2$ or the SU(3)$_1$ Wess-Zumino-Witten model.
\end{abstract}

\maketitle \emph{Introduction.}---Quantum entanglement has become a key concept in contemporary condensed matter physics. This is due in part to its ability to probe intrinsic topological order~\cite{PhysRevB.40.7387,PhysRevLett.96.110404,PhysRevLett.96.110405}.  Consider a density matrix, $\rho$, represented by the projector onto a many-body ground state. If the associated Hilbert space is partitioned into two non-overlapping regions $A$ and $B$, the entanglement between the regions A and B can be quantified either through the entanglement entropy (EE) or the entanglement spectrum (ES) obtained by the reduced density matrix $\rho_{\A}=\text{Tr}_{\B}\rho$. The EE is given by $S_\A=-\mathrm{Tr}(\rho_\A\mathrm{ln}\rho_\A)$, and the ES is defined as the spectrum of the entanglement Hamiltonian $H_{\text{E}}=-\text{ln} \rho_\A$~\cite{PhysRevLett.101.010504}. By definition, EE and ES depend on the chosen basis to partition (cut) the many-body Hilbert space. To resolve bulk and edge features of topological order, some form of spatial cut~\cite{PhysRevLett.101.010504,PhysRevLett.104.180502,PhysRevLett.104.156404,PhysRevLett.108.196402,PhysRevB.88.245137} (along with a particle cut~\cite{PhysRevLett.98.060401,PhysRevLett.106.100405}) is the predominantly used choice. This works well in systems with a bulk energy gap and hence an associated length scale.  Upon partitioning, the ES then mimics the spectral features located along the cut of an edge termination \cite{PhysRevLett.108.196402}. In particular, a set of universal entanglement modes related to the edge can be identified as distinct from generic entanglement weight through the entanglement gap (EG), which can be employed to define topological adiabaticity in the entanglement spectral flow~\cite{PhysRevLett.104.180502}. The spatial EG evolves in a way similar to the physical bulk gap of the topologically ordered phase, even though bulk gap closures occur at points of parameter space different from EG closures due to the unit fictitious temperature in $H_{\text{E}}$~\cite{PhysRevLett.113.060501}.

In order to understand the universal properties of entanglement in
critical systems, a spatial cut is not always a preferable choice~\cite{PhysRevA.78.032329,2013arXiv1303.0741L,1742-5468-2015-7-P07017}. Due to the absence of an energy gap, there will not be an appreciable concentration of entanglement weight localized along the cut. Furthermore, for geometries where a spatial cut induces multiple edges, such as a ring or torus, the entanglement modes couple between the edges, and complicate the resolution of individual modes. Instead, a momentum basis appears promising to detect universal critical entanglement profiles. The momentum-space ES was first introduced for spin-1/2 chains \cite{PhysRevLett.105.116805}. There, the notion of momentum relates to the Fourier transform of individual spin flip operators, and the total spin flip momentum, $M_\A$, of spin flips in momentum region A provides an approximate quantum number of $\rho_\A$. The spin fluid phase around the Heisenberg spin chain was found to exhibit a large EG and a U(1) counting of entanglement levels.  The momentum-space ES has been subsequently explored in the XXZ spin-1/2 chain \cite{PhysRevLett.113.256404}, spin-ladders \cite{PhysRevB.86.224422,2014arXiv1412.8612L}, and disordered systems \cite{PhysRevLett.110.046806,2014JSMTE..07..022A}. As an overarching principle, the momentum EG, along with the {\it universal} entanglement weight below it, require an interpretation different from the spatial cut. In a finite spin chain with no length scale, except for the UV lattice cutoff $1/a$ and IR chain length cutoff $1/L$, the EG cannot be directly related to a microscopic scale.
%{\color{red} Ronny, how do you know there is no scale associated with it?  I would have thought it somehow scales with one of the correction terms to the CFT --- a log or whatever.}
As a central conjecture emerging from previous work, the EG separates the non-universal part of the ES above it from the
universal part below it, which is determined by the 
associated conformal field theory.  We refer to this assumption  
as the universal bulk entanglement conjecture (UBEC).

In this Letter, we elevate this conjecture to a general principle, as we confirm it for several critical spin chains with different, intricate field theories. In particular, we analyze the momentum-space ES of several critical spin-1 chains, including the Takhtajan-Babujian (TB) point~\cite{Takhtajan1982479,Babujian1982479} and the Uimin-Lai-Sutherland (ULS) point~\cite{Uimin,Lai,Sutherland} associated with an SU$(2)_2$ and an SU$(3)_1$ Wess-Zumino-Witten (WZW) field theory, respectively. Both are critical points in the bilinear-biquadratic spin-1 model. (For a detailed complementary study of the real-space ES see Ref.~\onlinecite{1742-5468-2015-7-P07017}.) We pursue our analysis in two steps. First, we identify fine-tuned models related to SU$(2)_2$ and SU$(3)_1$ WZWs, which exhibit an infinite EG, relating to an extensive multiplicity of the eigenvalue zero in $\rho_A$. For SU$(3)_1$ WZW, this is the SU(3) symmetric generalization of the Haldane-Shastry model~\cite{PhysRevLett.60.635,PhysRevLett.60.639,PhysRevB.46.3191}. For SU$(2)_2$ WZW, this is the Pfaffian spin chain~\cite{Greiter,PhysRevB.85.195149}. Second, we turn to the TB and ULS point, where we find a finite EG along with a precise matching of energy levels for the universal entanglement content as compared to their associated infinite-EG models.

%%%%%%%%%%%%%%%%%%%%%%
{\em SU$(3)_1$ WZW theory.}---Starting from the spin-1/2 fluid phase
where the universal behavior we advocate was first observed for an
SU$(2)_1$ WZW theory, one way of generalization is the enlargement of
the internal symmetry group. The low energy sector of the ULS model is
described by SU$(3)_1$ WZW theory with central charge
$c=2$. Equivalently, SU$(3)_1$ WZW can be thought of as two gapless
free bosonic field theories, each with unit central
charge~\cite{PhysRevB.92.075128}. In terms of $S=1$ spin operators,
the Hamiltonian is given by
$H_{\text{ULS}}=\sum\limits_{\alpha=1}^N\bs{S}_{\alpha}\bs{S}_{\alpha+1}+\sum\limits_{\alpha=1}^N(\bs{S}_{\alpha}\bs{S}_{\alpha+1})^2$.
Periodic boundary conditions (PBCs) are implemented by placing the
sites on a unit circle embedded into the complex plane, with site
coordinates $\eta_\alpha=\exp(i 2\pi \alpha /N)$, $\alpha \in
\{1,\dots, N\}$.  Due to its enlarged symmetry, the ULS model can be
recast (up to a constant) in terms of SU(3) spin
vectors~\cite{1742-5468-2015-7-P07017}, 
\begin{align}
  H_{\text{ULS}}=\sum_{\alpha=1}^N\bs{J}_{\alpha}\cdot\bs{J}_{\alpha+1},
  \label{SU3_Heis} 
\end{align} 
where $\bs{J}_{\alpha}=\frac{1}{2}\sum_{\sigma\tau}c^{\dagger}_{\alpha
  \sigma}\bs{\lambda}^{\phantom{\dagger}}_{\sigma\tau}c^{\phantom{\dagger}}_{\alpha
  \tau}$ denotes the SU(3) spin vector on site $\alpha$ consisting of
the eight Gell-Mann matrices, and $\tau, \sigma \in \{\r,\g,\bl\}$. We
contrast model~\eqref{SU3_Heis} with the SU(3) Haldane-Shastry model
\begin{align}
  H_{\text{HS}}^{\text{SU(3)}}=\frac{2\pi^2}{N^2}
 \sum_{\alpha\ne\beta}^N
 \frac{\bs{J}_{\alpha}\cdot\bs{J}_{\beta}}{|\eta_\alpha-\eta_\beta|^2},
\label{SU3_HS}
\end{align}
where $|\eta_\alpha-\eta_\beta|$ is the chord distance along the ring.

In order to perform a momentum cut for the finite size ground state
of~\eqref{SU3_Heis} and~\eqref{SU3_HS}, we first need to specify the
operators which span the Hilbert space of the spin chain. In analogy
to the spin flip operators, $S_\alpha^+$, $S_\alpha^-$, which are
formed by the adjoint representation of SU(2), we have the color flip
operators $e^{\sigma \tau}_\alpha=c_{\alpha \sigma}^\dagger c_{\alpha
  \tau}^{\phantom{\dagger}}$ for SU(3). Assuming $N=0\;\text{mod}\;3$,
the ground states of~\eqref{SU3_Heis} and~\eqref{SU3_HS} will be SU(3)
singlets due to a generalized interpretation of the Marshall
theorem~\cite{marshall}. We write \begin{align}
  \ket{\psi_{0}}
  =\sum_{\{z;w\}}\psi_{0}[z;w] e^{\bl\g}_{z_1}\dots e^{\bl\g}_{z_{N/3}}
  e^{\r\g}_{w_1}\dots e^{\r\g}_{w_{N/3}}
  \ket{0_{\g}},\nonumber\\[-24pt]
  \phantom{m}\label{SU3_WF}
\end{align}
where the sum extends over all possible ways of distributing the positions $[z]\equiv z_1,\dots z_{N/3}$ of the blue (and $[w]\equiv w_1,\dots w_{N/3}$ of the red) particles. $|0_{\g}\rangle=\prod_{\a=1}^N c_{\a\g}^\dagger\vac $ is a reference state consisting only of green particles, on which we act with the color flip operators $e^{\bl\g}_\alpha$ and $e^{\r\g}_\alpha$. We define the momentum space operators $\tilde{e}^{\bl\g}_p$ and $\tilde{e}^{\r\g}_q$ 
\begin{align}
  e^{\bl\g}_{\a}
  =\frac{1}{\sqrt{N}}\sum_{p=1}^N \eab^{\, p}\,\tilde{e}^{\bl\g}_p,
  \quad 
   e^{\r\g}_{\b}
   =\frac{1}{\sqrt{N}}\sum_{q=1}^N \ebb^{\, q}\,\tilde{e}^{\r\g}_q,
%
%  e^{\tau\sigma}_{\a}
%  =\frac{1}{\sqrt{N}}\sum_{p=1}^N\eab^{\, p}\,\tilde{e}^{\tau\sigma}_p,
%  \quad 
%   \tilde{e}^{\tau\sigma}_p 
%   =\frac{1}{\sqrt{N}}\sum_{\a=1}^N\ea^{\, p}\,e^{\tau\sigma}_{\a},
  \label{SU3FT}
\end{align}
where $p,q\in\{1,\dots, N\}$ are integer spaced momentum indices.  Substitution of \eqref{SU3FT} into \eqref{SU3_WF} yields
\begin{align}
  \ket{\psi_0}
  =\sum_{\{p;q\}}\tilde\psi_0[p;q]\,
  \tilde{e}^{\bl\g}_{p_1}\dots\tilde{e}^{\bl\g}_{p_{N/3}}
  \tilde{e}^{\r\g}_{q_1}\dots \tilde{e}^{\r\g}_{q_{N/3}}
  \ket{0_\g},%\nonumber
  \label{SU3ketpsitilde}
\end{align}
\begin{align}
  \tilde{\psi}_0[p;q] = \sum_{\{z;w\}}\!\psi_0[z;w]\;
  \zb_1^{p_1}\dots\zb_{N/3}^{p_{N/3}}\wb_1^{q_1}\dots\wb_{N/3}^{q_{N/3}}.
  \label{SU3psitilde}
%  \nonumber
\end{align} 
Note that while there trivially is a hard-core constraint for the color flip operators in real space, there is no such condition in momentum space.  This significantly enlarges the number of basis states.  For our purposes, it is best to write the ground state in a momentum space occupation number basis, \begin{align}
  \label{mom_coeff}
  \ket{\psi_0}
  =\sum_{\{n;m\}} \tilde\phi_0[n;m] 
  \ket{n_1,\ldots,n_N;m_1,\ldots,m_N}, 
\end{align} 
where $n_p$ ($m_q$) is the number of times momentum index $p$ ($q$) for color flips from green to blue (red) appears in \eqref{SU3ketpsitilde}. The ket in \eqref{mom_coeff} is hence given by \begin{align}
%  \ket{n_1,\ldots,n_N;m_1,\ldots,m_N}=
  \ket{n_1\ldots;m_1,\ldots}=
   \prod_{p=1}^N 
   \frac{(\tilde{e}_{p}^{\bl\g})^{n_p}}{\sqrt{n_p!}}  
   \prod_{q=1}^N 
    \frac{(\tilde{e}_{q}^{\r\g})^{m_q}}{\sqrt{m_q!}}  
%   \prod_{p=q=1}^N 
%   \frac{(\tilde{e}_{p}^{\bl\g})^{n_p}(\tilde{e}_{p}^{\r\g})^{m_p}}
%   {\sqrt{n_p!\,m_p!}}
   \ket{0_g}.
  \label{eq:basis}
\end{align}
We arrive at Eq.~\eqref{mom_coeff} after obtaining the real space ground state via exact diagonalization. Due to the exponential numerical cost of the many-particle Fourier transform, the maximal size we are able to reach is $N=15$.

%\vspace{50pt}
% With $n_{\tau, l} \equiv n_{\tau, l} [\{k\}, \{m\}] \in \mathbb{N}_0$ encoding the occupation number of particles with color $\tau$ and momentum $l$, the Fock basis yields 
% \begin{align}
%   | n_b ; n_r \rangle:=\prod_{l=1}^N
%   \frac{(\tilde{e}_{l}^{bg})^{n_{b,l}}(\tilde{e}_{l}^{rg})^{n_{r,l}}}
%   {\sqrt{n_{b,l}!n_{r,l}!}}|0_g\rangle,%\nonumber
% \end{align}
% and the momentum-space wave function gives
% \begin{align}
%   |\psi_0\rangle
%   =\sum_{n_b;n_r}\frac{\tilde\psi_0[\{k\};\{m\}]}
%   {\prod_{l=1}^N \sqrt{n_{b,l}!n_{r,l}!}}| n_b; n_r \rangle,%\nonumber
% \label{mom_coeff}
% \end{align}
% where the sum runs over all distinct momentum occupation partitions of the blue and red particles. We arrive at Eq.~\eqref{mom_coeff} after obtaining the real space ground state via exact diagonalization. Due to the exponential numerical cost of the many-particle Fourier transform, the maximal size we are able to reach is $N=15$.

\begin{figure*}[t!]
\subfloat[][SU(3) Haldane-Shastry Model]{
 \includegraphics[width=.4\linewidth]{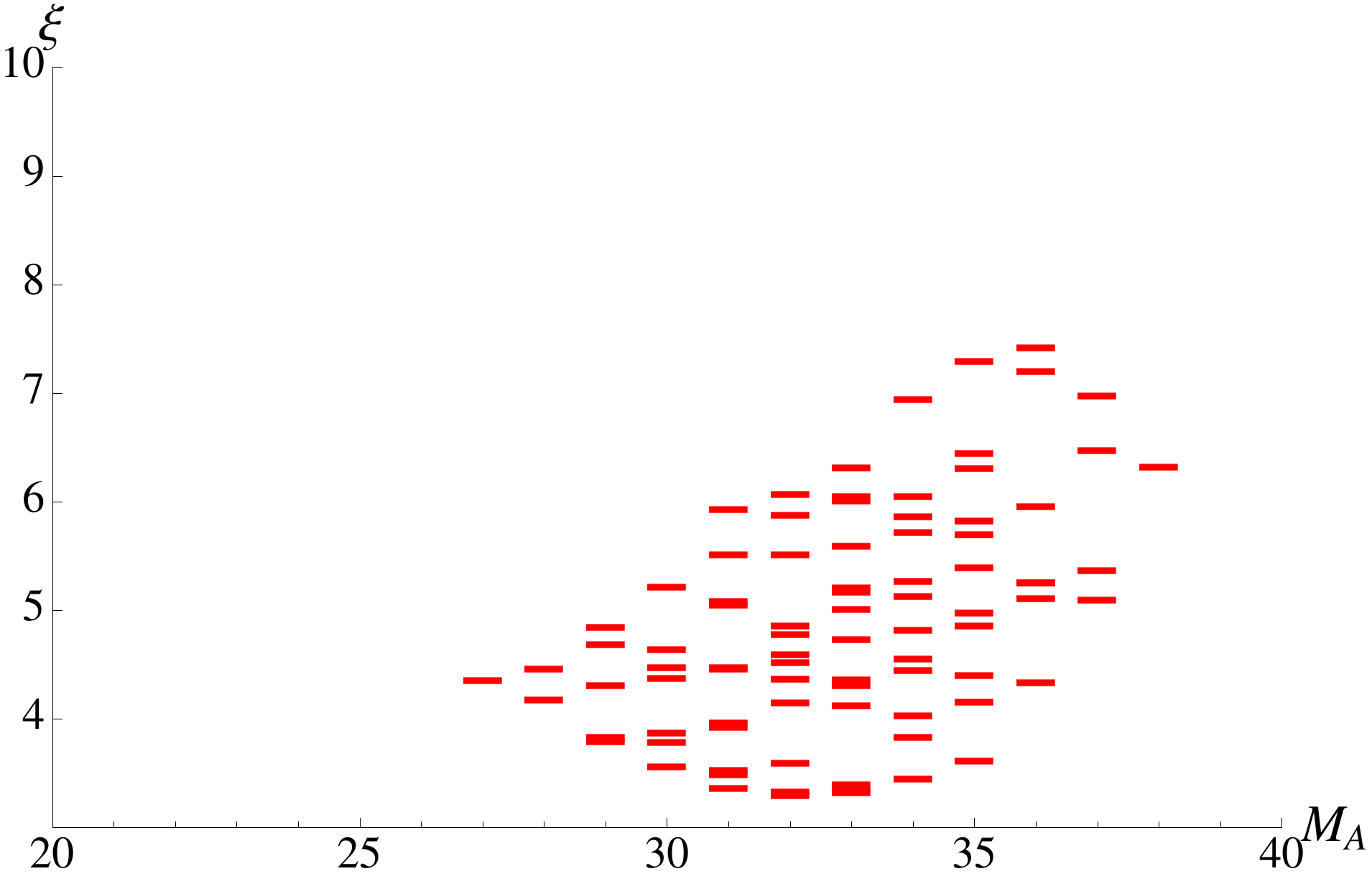}
\label{HS_ES_15_6}
}
\subfloat[][Uimin-Lai-Sutherland Point]{
 \includegraphics[width=.4\linewidth]{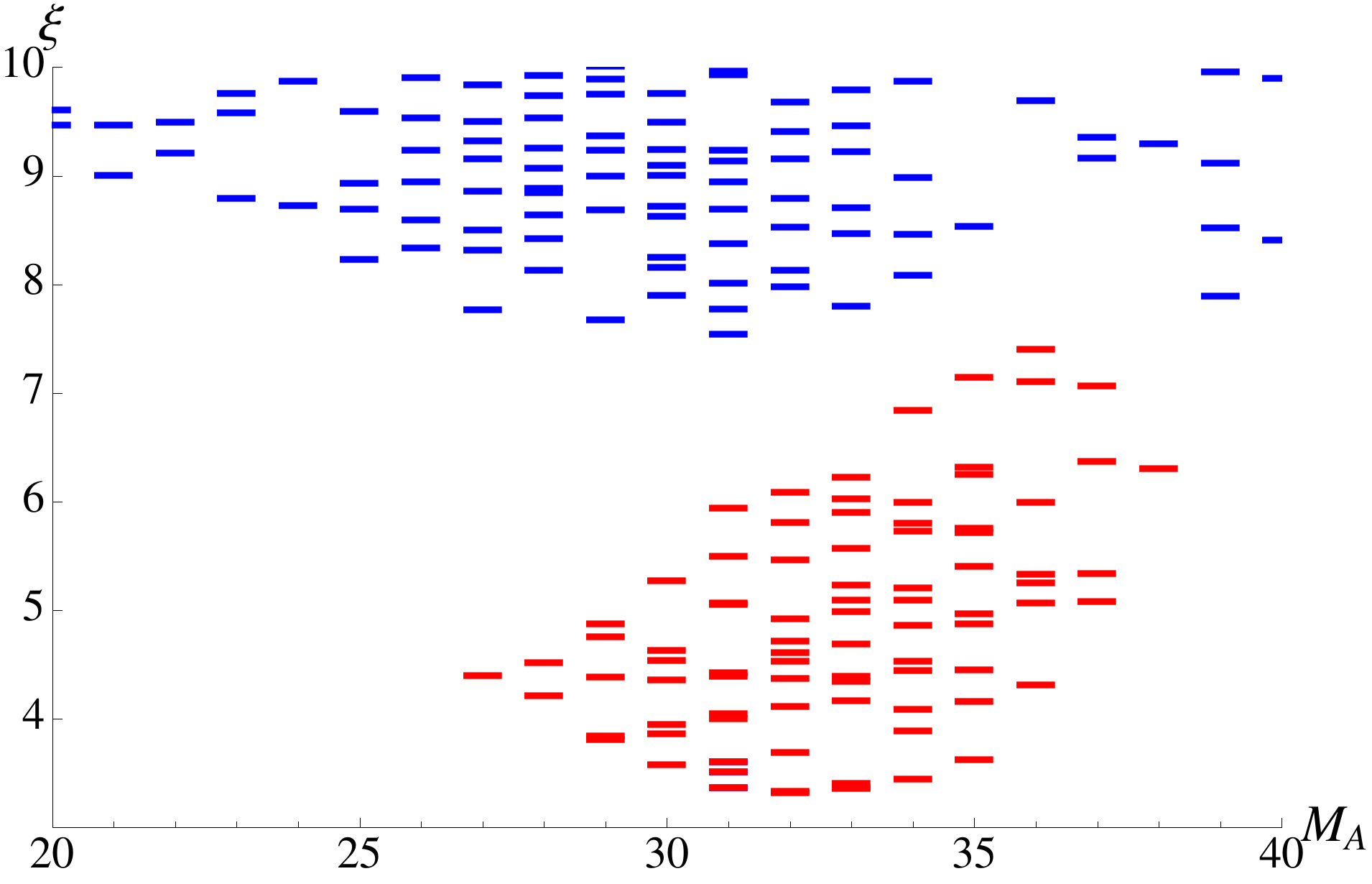}
\label{HEIS_ES_15_6}
\newline
}
\caption{(color online) (a) ES of Eq.~\eqref{SU3_HS} and (b) ES of the ULS point for $(N,N_A)=(15,6)$, $3$ red and $3$ blue particles. 
% In both cases, $N=15$ and $N_A=6$ (3 red-green colorflip particles and 3 blue-green colorflip particles). 
At the ULS point, generic entanglement levels (blue) are separated by a finite EG from the universal entanglement levels (red). The eigenvalues, $\xi$, are plotted versus the total momentum of region $A$.  Throughout this work, $\rho_A$ is normalized such that $\tr\rho_A=1$ for each $N_A$. The universal entanglement in (a) and (b) matches the counting of SU(3)$_1$ WZW theory, supporting the UBEC.}
\label{15sites}
\end{figure*} We are now prepared to calculate the momentum ES for~\eqref{SU3_Heis} and~\eqref{SU3_HS}. Assuming $N$ odd, we partition momentum into regimes 
$\A
=\{p\,|\,p\leq\frac{N+1}{2}\}\otimes\{q\,|\,q\leq\frac{N+1}{2}\}$ and 
$\B=\{p\,|\,p>\frac{N+1}{2}\}\otimes\{q\,|\,q>\frac{N+1}{2}\}$. Region A and B are decomposed in terms of total momentum, $M=M_\A+M_\B$, and particle number, $N=N_\A+N_\B$, which are given by $M_{\A/\B}=\sum_{p\in \A/\B} n_{p}p + \sum_{q\in \A/\B} m_{q}q$ and 
$N_{\A/\B}=\sum_{p\in \A/\B} n_{p}  + \sum_{q\in \A/\B} m_{q} $. 
The crystal momentum is given by $M_{\A/\B}^{\text{c}}=M_{\A/\B}\,\text{mod}\,N$, and is always an exact quantum number of $\rho_{\A/\B}$. In general, however, even $M_{\A/\B}$ is a good approximate quantum number. For an $N=12$ ground state of~\eqref{SU3_Heis}, more than 99\% of the total amplitude resides in the $M=\frac{N^2}{3}$ sector and less than 1\% in all other sectors. It is a central observation that $M$ (and $M_{\A/\B}$) is a good approximate quantum number as long as the internal spin symmetry is unbroken or only weakly broken~\cite{PhysRevLett.105.116805,PhysRevLett.113.256404}.

The ground state of~\eqref{SU3_HS} retains $M_{\A/\B}$ as an {\it exact} quantum number. Fig.~\ref{HS_ES_15_6} displays its $(N,N_\A)=(15,6)$, $N_{\A,\r}=N_{\A,\bl}=3$ sector of $H_{\text{E}}$ with spectral levels denoted by $\xi$. We observe a large degeneracy of entanglement levels at infinity, corresponding to eigenvalues zero of $\rho_A$.  The counting $1,2,5$ of the ES levels from left to right matches the state counting of two gapless U(1) bosons until we reach a finite size limit. All properties above are understood on analytic footing:
For~\eqref{SU3_HS}, $\psi^{\text{HS}}_{0}[z;w]$ is given by~\cite{PhysRevB.73.235105}
\begin{align}
\psi^{\text{HS}}_{0}[z;w]=\prod_{i<j}^{N/3}(z_i&-z_j)^2(w_i-w_j)^2
\nonumber\\[-10pt]
&\cdot\prod_{i,j=1}^{N/3}(z_i-w_j)\prod_{i=1}^{N/3}z_i w_i.
\label{HS_WF}
\end{align}
Note that one can write the ground state in terms of color flip operators for any pair of colors (up to a minus sign) \cite{PhysRevB.75.024405}. By virtue of a momentum-conserving orbital squeezing relation between Fock states of non-zero weight, Eq.~\eqref{HS_WF} has all of its weight in the sector $M=N^2/3$. To understand this, note that \eqref{HS_WF}, in its polynomial form, is equivalent to the spin-singlet bosonic Halperin-(221) fractional quantum Hall (FQH) state \cite{halperin1983theory} with filling fraction $\nu=\frac{2}{3}$. Vice versa,
%this is consistent with the finding that 
the bosonic Halperin-(221) state exhibits SU(3) symmetry~\cite{PhysRevLett.82.5096}.
As the Halperin-(221) state obeys certain squeezing properties \cite{PhysRevB.84.205134,PhysRevB.84.045127}, so does~\eqref{HS_WF}. In terms of critical theories,~\eqref{SU3_HS} is special in the sense that the finite size ground state does not contain corrections as compared to the thermodynamic field theoretical content of entanglement.  

Turning to the ES at the ULS point~\eqref{SU3_Heis} in Fig.~\ref{HEIS_ES_15_6}, we observe an EG present for all $M_A$, which separates the non-universal components at higher $\xi$ from universal levels which match with the entanglement levels of~\eqref{SU3_HS}. As one increases the system size, the relative importance of non-universal entanglement levels would decrease while the universal entanglement weight becomes successively dominant and stays separated from non-universal levels through the EG. It implies that the UBEC also holds for critical spin chains described by SU$(3)_1$ WZW theory.

\begin{figure*}[t!]
\subfloat[][Pfaffian Spin Chain]{
 \includegraphics[width=.33\linewidth]{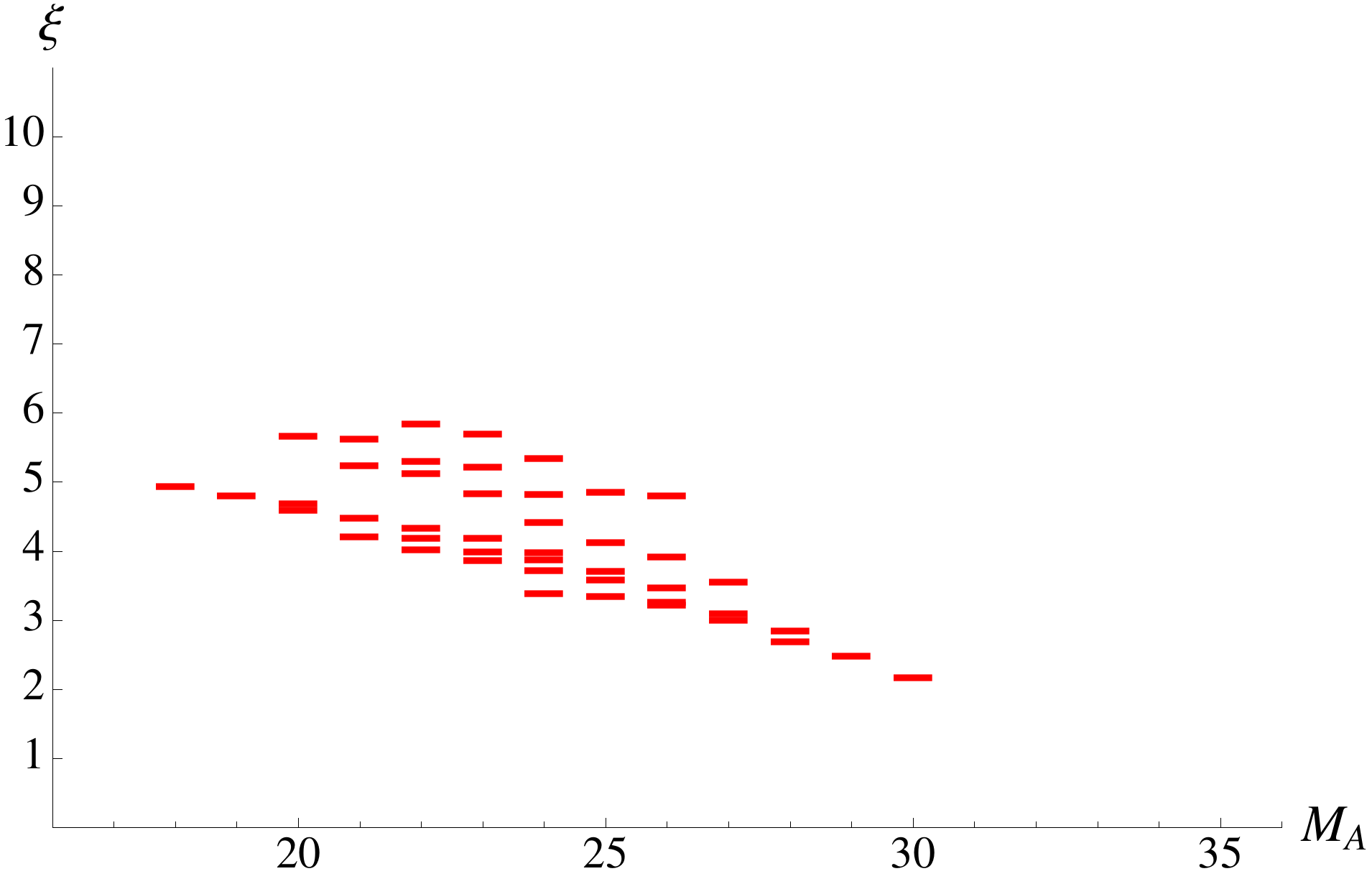}
\label{Pfaff_ES}
}
\subfloat[][Takhtajan-Babujian Point]{
 \includegraphics[width=.33\linewidth]{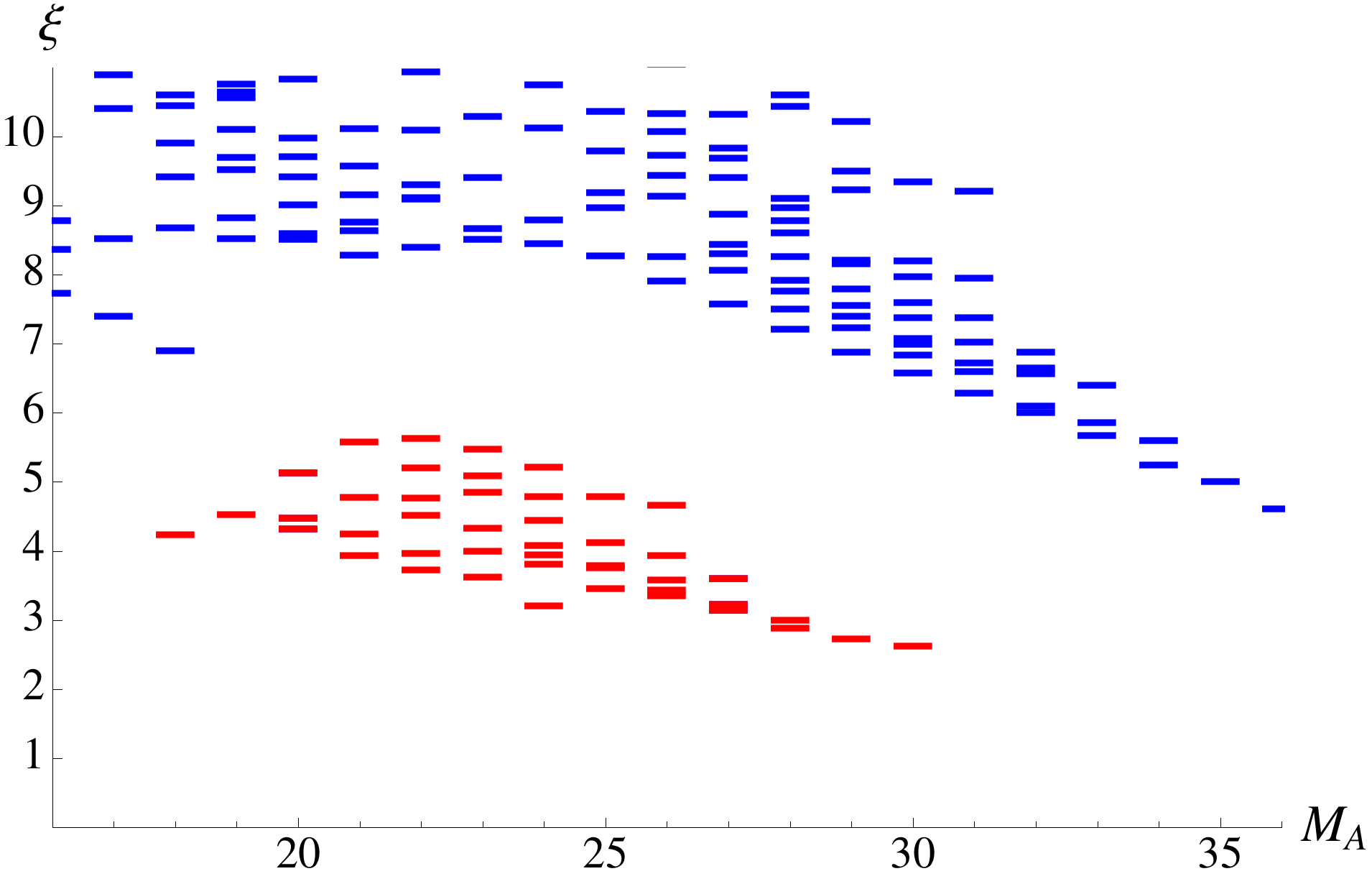}
\label{TB_ES}
}
\subfloat[][$J_1-J_3$ Model]{
 \includegraphics[width=.33\linewidth]{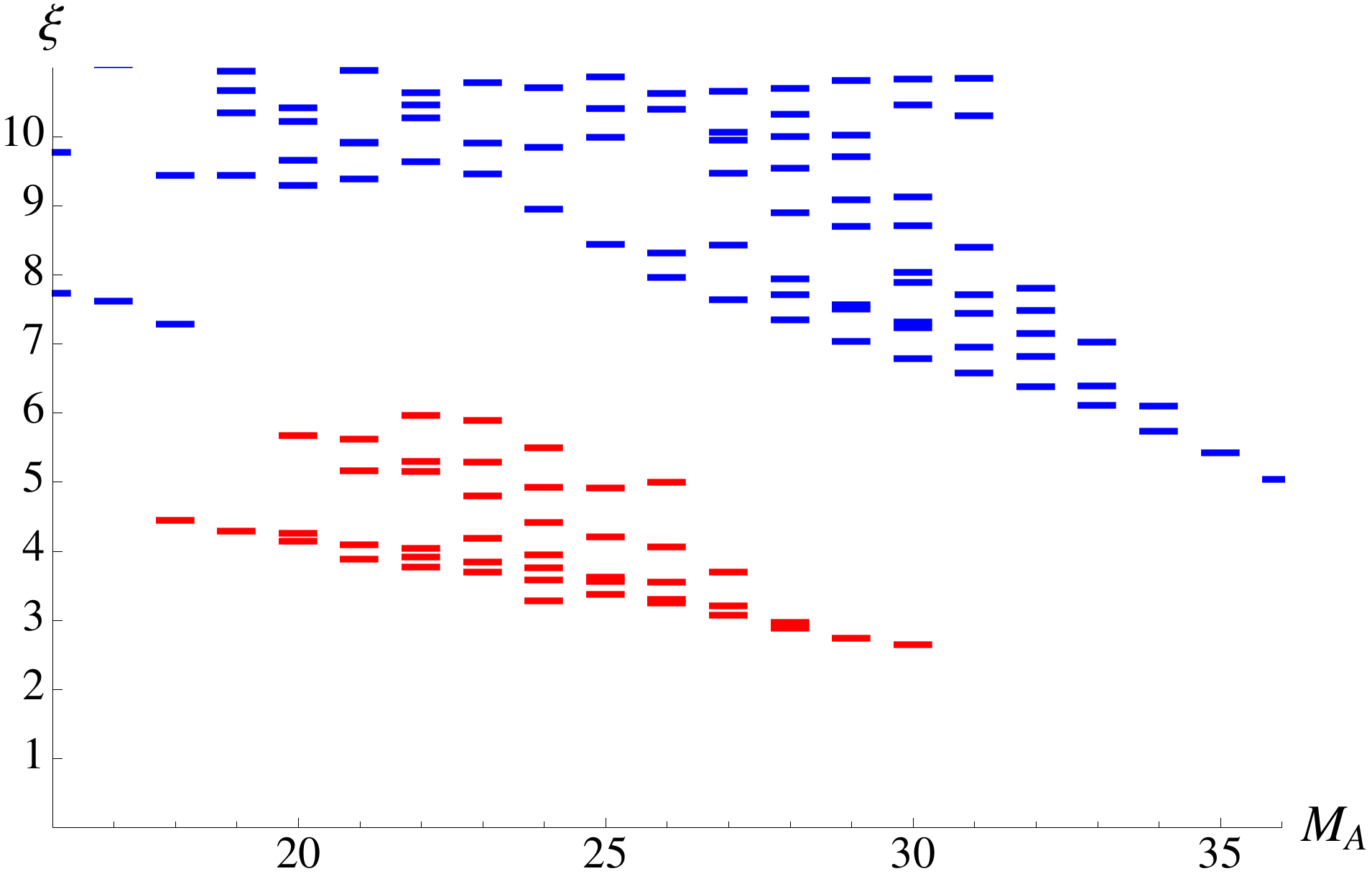}
\label{J1J3_ES}
}
\caption{(color online) (a) ES of Eq.~\eqref{MARTIN}, (b) ES of the TB point, and (c) ES of the $J_1-J_3$ model for $(N,N_A)=(12,6)$. 
%In all three cases, $N=12$ and $N_A=6$. 
  In (b) and (c), the universal (red) and non-universal (blue) entanglement levels are separated by a finite EG. For (a)-(c), the counting of the universal levels matches the counting of SU(2)$_2$ WZW theory, consistent with the UBEC.}
\label{FiguresE}
\end{figure*}

%%%%%%%%%%%%%%%%%%%%%%%
{\em SU$(2)_{k=2}$ WZW theory.}---Another way to explore the reach of the UBEC is the extension to higher level $k>1$ Wess-Zumino terms in the field theory description of critical spin chains. Higher $k$ links to multi-critical points which in general do not represent gapless spin fluid phases, but rather phase transition points~\cite{PhysRevB.36.5291}. For SU$(2)_2$ WZW theory, several model instances have been found for spin-1 chains such as the TB spin chain $H_{\text{TB}}=\sum\limits_{\alpha=1}^N\bs{S}_{\alpha}\bs{S}_{\alpha+1}-\sum\limits_{\alpha=1}^N(\bs{S}_{\alpha}\bs{S}_{\alpha+1})^2$. An analytic lattice realisation of SU$(2)_2$ WZW theory has been found for the Pfaffian spin chain~\cite{Greiter,1742-5468-2011-11-P11014}, 
\begin{align} 
\label{MARTIN}
 H^{\text{Pf}}&=\frac{2\pi^2}{N^2}\Bigg[\sum_{\alpha\ne\beta}^N
   \frac{\bs{S}_{\alpha}\bs{S}_{\beta}}{|\eta_\alpha-\eta_\beta|^2}\\\nonumber
   &-\frac{1}{20}\sum_{\substack{\alpha,\beta,\gamma\\ \alpha\neq\beta,\gamma}}^N
   \frac{(\bs{S}_{\alpha}\bs{S}_{\beta})(\bs{S}_{\alpha}\bs{S}_{\gamma})
     +(\bs{S}_{\alpha}\bs{S}_{\gamma})(\bs{S}_{\alpha}\bs{S}_{\beta})}
   {(\bar{\eta}_{\alpha}-\bar{\eta}_{\beta})(\eta_{\alpha}-\eta_{\gamma})}\Bigg].
\end{align}
The low-energy theory is described by a massless bosonic and Majorana field consistent with $c=1+0.5$~\cite{PhysRevB.85.195149}. 
Numerical evidence for SU$(2)_2$ critical behaviour has been
idependently 
found for a truncated version of \eqref{MARTIN},
%$H^{J_1\text{-}J_3}=
%\sum\limits_{\alpha=1}^N J_1\bs{S}_{\alpha}\cdot \bs{S}_{\alpha+1} 
%+ J_3[(\bs{S}_{\alpha-1}\cdot \bs{S}_{\alpha})(\bs{S}_{\alpha}\cdot 
%\bs{S}_{\alpha+1})+\text{h.c.}]$
\begin{align}
  H^{J_1\text{-}J_3}\hspace{-1pt}&=\hspace{-1pt}
  \Bigg[\sum\limits_{\alpha=1}^N 
  \bs{S}_{\a}\bs{S}_{\a+1} \hspace{-1pt}\\\nonumber
  &+\hspace{-1pt}\frac{J_3}{J_1}\bigl[
    (\bs{S}_{\a-1}\bs{S}_{\a})(\bs{S}_{\a}\bs{S}_{\a+1})+\text{h.c.}
    \bigr]\Bigg],\label{eq:HJ1J3}
\end{align}
at $J_3/J_1 \approx 0.11$,
with a central charge $c=1.5$~\cite{PhysRevLett.108.127202}.

The singlet ground state of any spin-1 chain of length $N$ ($N$ even) reads
\begin{equation}
|\psi_0^{S=1}\rangle=\sum_{\{z\}}\psi_0(z_{1},\dots, z_{N})\, \tilde{S}_{z_1}^{+}\dots\tilde{S}_{z_N}^{+}|-1\rangle_N,
\label{WF_1}
\end{equation}
where the sum extends over all possible configurations of $N$ spin-flip operators (allowing for at most two spin flips on the same site),  $|-1\rangle_N=\otimes^N_{i=1}|1,-1\rangle$ is the vacuum with all spins in the $s^z=-1$ state, and 
$\tilde{S}^{+}_{\alpha}=\frac{1}{2}(S^z_\alpha+1)S^{+}_\alpha$ is a renormalized spin flip operator~\cite{PhysRevLett.102.207203}. They are the natural choice to unify a Pfaffian polynomial of spin flip coordinates with the singlet property of the resulting wave function, such that the ground state of~\eqref{MARTIN} yields
\begin{equation}
\psi^{\text{Pf}}_0(z_1,\dots,z_N)=
\mathrm{Pf}\bigg(\frac{1}{z_i-z_j}\bigg)
\prod\limits_{i<j}^N(z_i-z_j)\prod\limits_{i=1}^N z_i,
\label{Pfaff}
\end{equation}
where $\text{Pf}(1/z_i-z_j)=\mathcal{A}[(1/(z_1-z_2) \dots 1/(z_{N-1}-z_N)]$. An alternative construction of~\eqref{Pfaff} is given by  the symmetrization over two S=1/2 Haldane-Shastry chain states~\cite{PhysRevLett.102.207203,PhysRevB.84.140404}.
We Fourier transform the spin-flip operators as
\begin{equation}
  S_\a^{+}
  =\frac{1}{\sqrt{N}}\sum_{q=1}^{N}\eab^q\, \tilde{S}_q^+,\quad
  \tilde{S}_q^{+}
  =\frac{1}{\sqrt{N}}\sum_{\a=1}^{N}\ea^q\, S_\a^+.
  \label{FT}
\end{equation}
%where $q\in\{1,\dots, N\}$ is the momentum index. 
Substituting~\eqref{FT} into~\eqref{WF_1}, we find
\begin{align}
  |\psi^{S=1}_0\rangle
  =\sum_{\{q\}}\tilde{\psi}_0(q_1,\dots, q_N)\,
  \tilde{S}_{q_1}^+\dots \tilde{S}_{q_N}^+\ket{-1}_N,
  \\
  \tilde{\psi}_0(q_1,\dots,q_N)
  =%\frac{1}{N^{\frac{N}{2}}}
  \sum_{\{z\}}\psi_0(z_{1},\dots,z_{N})\,\zb_1^{\,q_1}\dots\zb_N^{\,q_N}.
\end{align}
From the Fourier transformed ground state, we obtain the momentum-space ES. We partition our system in two regions, A and B, by dividing the momentum-space occupation basis as $\A =\{q\,|\,q<\frac{N}{2}\}$ and $\B= \{q\,|\,q>\frac{N}{2}\}$. Each region is decomposed in terms of number of particles $N=N_\A+N_\B=\sum_{q=1}^N n_q$ and total momentum $M=M_\A+M_\B=\sum_{q=1}^N n_qq$, where $n_q$ denotes the occupation number of a given momentum $q$. As previously seen for the SU(3) case, $M^c=M\;\text{mod}\;N$ is an exact quantum number, while $M$ in general is not. By virtue of being a squeezing state, however, \eqref{Pfaff} has all of its weight in the sector $M=N^2/2$. Similarly, it turns out that $M=N^2/2$ is the strongly preferred sector for the TB model and the $J_1-J_3$ model as well, rendering $M_A$ a good approximate quantum number. (For instance, the $N=10$ TB ground state has $94 \%$ of its total weight in the $M=50$ sector.)
%   a good quantum number, except at certain points but if the ground state has most of its weight in one momentum sector, it is still a useful approximate quantum number. The TB point has $94.6$ percent of its weight in the $\frac{L^2}{2}$ momentum sector for $N=10$. We thus consider the total momentum to be an approximately good quantum number.

Fig.~\ref{Pfaff_ES} depicts the $(N,N_\A)=(12,6)$ ES of~\eqref{Pfaff}
in comparison to the ES of the TB and $J_1$-$J_3$ ground state in
Fig.~\ref{TB_ES} and Fig.~\ref{J1J3_ES}, respectively. For all ES, we
observe a matching of universal levels which corresponds to counting
$1,1,3,\dots$ of the low-lying entanglement levels from left to
right. This corresponds to the energy levels of a boson and a Majorana
fermion with anti-periodic boundary
conditions~\cite{PhysRevB.53.13559}. For $N_\A=7$ not shown, the
observed counting is $1,2,4,\dots,$ and as such also consistent with
the previous finding \footnote{The finite effects in the ES counting
  as well as the perspective from root partition monomials in momentum
  space show that
  the $(N,N_A)=(12,6)$ sector corresponds to the $\mathbb{1}$ branch
  and the $(N,N_A)=(12,7)$ sector to the $\Psi$ branch. The $\sigma$
  branch is resolved by the ES analysis related to the ground state
  of~\eqref{MARTIN} for $N$ odd~\cite{PhysRevB.85.195149}.}. While
Fig.~\ref{FiguresE}a shows no non-universal entanglement weight beyond
the universal levels, \ie an extensive number of zero modes in
$\rho_A$. This is due to the monomial equivalence
between~\eqref{Pfaff} and the bosonic Moore-Read
state~\cite{Moore1991362}. Fig.~\ref{FiguresE}b and
Fig.~\ref{FiguresE}c exhibit different non-universal entanglement
weight, which is again separated from the universal weight by an EG,
in agreement with the UBEC. Note that in analytically unresolved cases
such as the model $J_3/J_1\approx 0.11$ in $H_{J_1-J_3}$, the momentum
entanglement fingerprint provides a particularly elegant tool to identify the critical theory.

{\em Conclusions.}---At the example of critical spin-1 chains, we have
provided evidence that the universal bulk entanglement
conjecture for critical spin chains generically holds for SU(N)$_k$
Wess-Zumino-Witten theories. From a broader perspective, our work
highlights that entanglement spectra do not only provide universal
fingerprints for topological phases, but also for critical systems. 
%{\color{blue} Our finding prompts future research, including the possible classification of phase transitions between topological phases from the viewpoint of momentum-space entanglement spectra.} \sout{Many directions of future research unfold from this finding, including a classification of phase transitions between topological phases from the viewpoint of entanglement.} 

%%%%%%%%%%%%%%%%%%%%%%%%%%%%%%%%%%%%%%%%%%%%%%%%%
{\em Acknowledgments.}---%R.L. thanks N. Regnault for helpful discussions and pointing out Ref.~\cite{PhysRevLett.82.5096}. 
R.L. was supported by National Science Foundation (NSF) Graduate Research Fellowship award number 2012115499, a fellowship from the Office of Graduate Studies at The University of Texas at Austin, and NSF Grant No. DMR-0955778. 
%R.L. thanks the hospitality of the Institute for Theoretical Physics, University of W\"{u}rzburg where the beginning stages of this work took place. 
P.L. and G.A.F. acknowledge financial support through NSF Grant
No. DMR-1507621. N.~R. was supported by the Princeton Global
Scholarship. M.G. and R.T. are supported by DFG-SFB 1170. R.T. was
supported by DFG-SPP 1458 and the ERC starting grant TOPOLECTRICS of the European Research Council (ERC-StG-Thomale-2013-336012). We acknowledge the Texas Advanced Computing Center (TACC) at The University of Texas at Austin for providing computing resources that have contributed to the research results reported within this paper. URL: http://www.tacc.utexas.edu
\bibliography{spin_1}

\end{document}